\documentclass{article}

\usepackage{arxiv}

\usepackage[utf8]{inputenc} 
\usepackage[T1]{fontenc}    
\usepackage{hyperref}       
\usepackage{url}            
\usepackage{booktabs}       
\usepackage{amsfonts}       
\usepackage{nicefrac}       
\usepackage{microtype}      
\usepackage{lipsum}
\usepackage{subfig}
\usepackage{graphicx}
\usepackage{amsmath}
\usepackage[ruled,vlined,linesnumbered]{algorithm2e}
\usepackage{color}
\usepackage{enumitem}
\usepackage{adjustbox}

\SetCommentSty{mycommfont}

\title{Simplification of Indoor Space Footprints}

\author{
  Joon-Seok Kim \\
  Department of Geography and Geoinformation Science\\
  George Mason University\\
  Fairfax, VA 22030 \\
  \texttt{jkim258@gmu.edu} \\
  \AND
  Carola Wenk\thanks{This research was supported by National Science Foundation grant CCF 1637576.}\\
  Department of Computer Science\\
  Tulane University\\
  New Orleans, LA 70118 \\
  \texttt{cwenk@tulane.edu} \\
}

\begin{document}
\maketitle

\setlist[itemize]{leftmargin=*}

\begin{abstract}
Simplification is one of the fundamental operations used in geoinformation science (GIS) to reduce size or representation complexity of geometric objects. Although different simplification methods can be applied depending on one's purpose, a simplification that many applications employ is designed to preserve their spatial properties after simplification. This article addresses one of the 2D simplification methods, especially working well on human-made structures such as 2D footprints of buildings and indoor spaces. The method simplifies polygons in an iterative manner. The simplification is segment-wise and takes account of intrusion, extrusion, offset, and corner portions of 2D structures preserving its dominant frame. 
\end{abstract}


\section{Introduction}

\begin{figure}[t]
	\centering
	\subfloat[initial\label{fig:intial}]{%
       \includegraphics[width=0.33\textwidth]{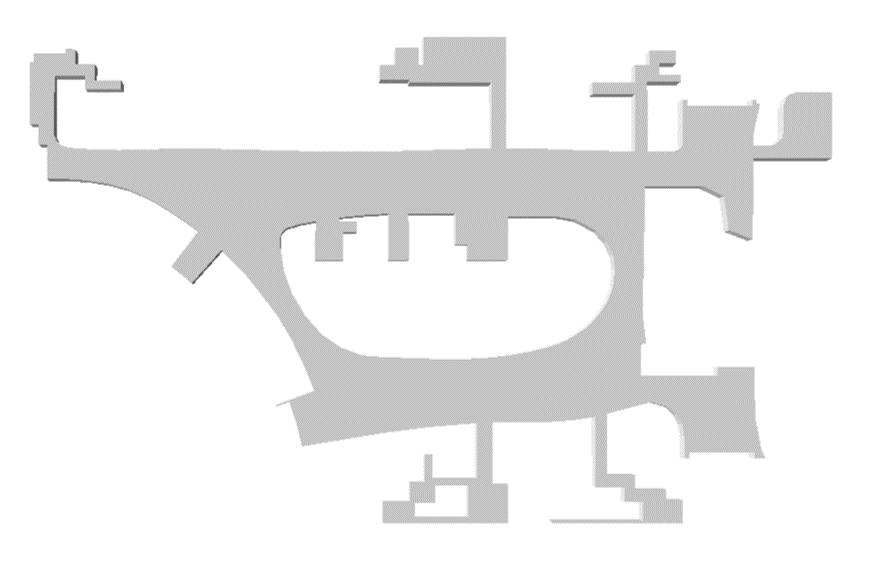}
    }
    \subfloat[Ramer-Douglas-Peucker algorithm \label{fig:dp-result}]{%
       \includegraphics[width=0.33\textwidth]{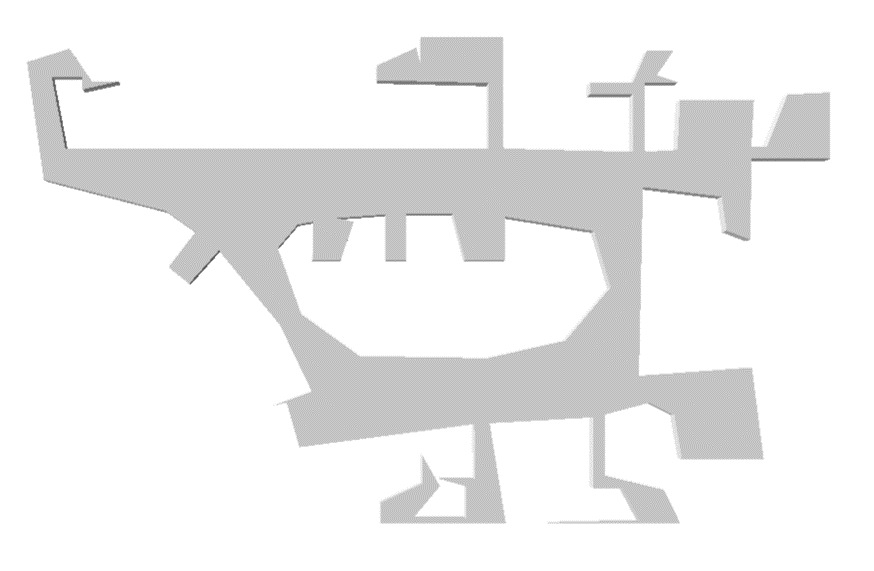}
    }
    \subfloat[simplification preserving spatial properties \label{fig:our-result}]{%
       \includegraphics[width=0.33\textwidth]{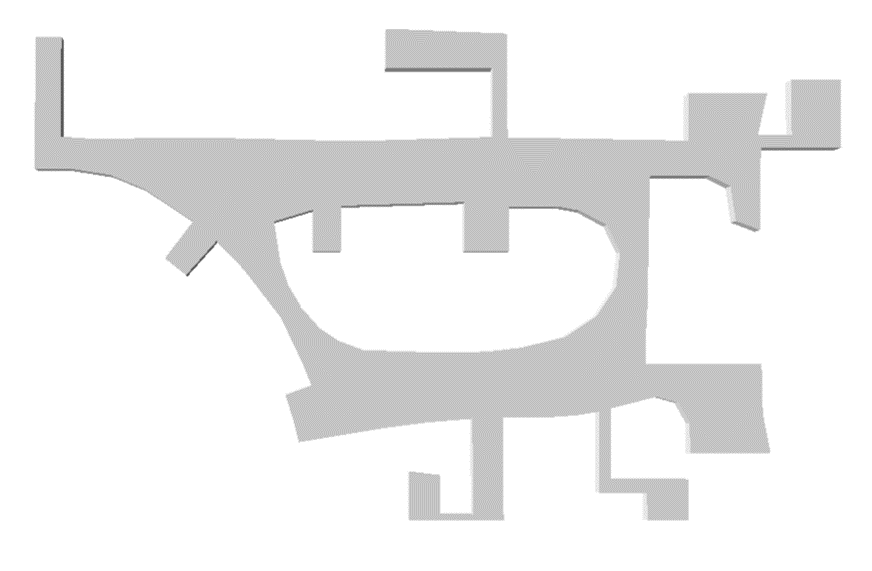}
    }
    \caption{Example of simplification of a complex indoor space: (b) and (c) have 103 and 102 segments, respectively\vspace{-0.2cm}}
    \label{fig:comparison}
\end{figure}

Simplification is one of the methods used for generating data at different levels of detail (LoDs) from precise data. The more compact size of simplified data is desirable in a variety of data processing tasks including data transmission. Let $P$ be a polygon, $P'$ be a simplified polygon, and $D(P,P')$ be the distance between $P$ and $P'$. The computational geometry community distinguishes between two variants of curve simplification: While the min–$\#$ problem is to find $P'$ with the minimum number of vertices such that $D(P,P') \leq \varepsilon$, the min-$\varepsilon$ problem is to find $P'$ of at most $k$ vertices such that $D(P,P')$ is minimized. Depending on the constraints on the location of vertices of $P'$, the problem can be categorized into (1) vertex-restricted, (2) curve-restricted, and (3) non-restricted simplification, see \cite{kklmw-gcs-19}. The Ramer-Douglas-Peucker (RDP) algorithm \cite{Douglas:1973,ramer1972iterative} is widely used in practice and provides a vertex-restricted approximation to simplify 2D polylines and polygons. However, it does not provide any quality guarantees and it may not preserve essential shape of the entire footprint because it does not reflect a specific form (see Figure \ref{fig:dp-result}). Although Figures \ref{fig:dp-result} and \ref{fig:our-result} are similar in terms of the number of segments, the figures demonstrate the RDP cannot preserve spatial features such as intrusion, extrusion, offset, and corners. In geographic information science (GIS), simplification is considered a type of generalization process \cite{weibel1999generalising}. These processes or operations consider not only metric constraints but also topological, semantic, and Gestalt constraints. In particular, Gestalt constraints are used to preserve the characteristics of spatial features such as a room.

This article presents a method that progressively simplifies 2D polygons by preserving their spatial properties in a iterative manner. The method was originally introduced by Kim and Li \cite{kim2019simplification} to simplify 3D indoor spaces (e.g., rooms and hallways) that can be represented with the prism model \cite{3dgeoinfo08:Kim08,fl-3gdds-78}, which is an alternative 3D data model. As shown in Figure \ref{fig:our-result}, the simplification can be applied to footprints of complex buildings such as shopping malls or subway stations. This article elaborates on the simplification of 2D footprints of \cite{kim2019simplification} so that anyone can implement and apply it to their applications. In the next section, the detailed method is described.

\section{Simplification of Polygons}

Let $P$ be a simple planar polygon without holes. $P$ can be represented by the circular sequence of $n$ line segments $\langle s_0,\ldots,s_{n-1} \rangle$ describing the boundary of $P$ in counterclockwise order, where $p_i\in\mathbb{R}^2$ and each pair of vertices $s_i=(p_i, p_{i+1})$ represents a line segment, for all  $0\leq i<n$. The sequence of vertices of $P$ is circular such that $p_{i}=p_{(i\mbox{ mod }n)}$ for $i \in \mathbb{Z}$. Let $\overline{s_i}$ be the length of $s_i$, and let $\widehat{s_{i}}$ be the internal angle between two consecutive segments $s_{i}$ and $s_{i+1}$, i.e., $\widehat{s_{i}}=\angle p_i p_{i+1} p_{i+2}$.  In the following, $Q$ denotes a priority queue containing line segments $s_i$ sorted by their length $\overline{s_i}$. Given a {\em distance threshold} $\tau$, an {\em angle threshold} $\varepsilon$, a {\em collinearity threshold} $\delta$, and a {\em joining distance threshold} $\gamma$, the simplification is designed to preserve the overall shape of a 2D polygon according to the following rules: 

\begin{itemize}
\item \emph{Rule-1}: The shorter the line segment, the smaller the effect on the overall shape of the geometry.
\item \emph{Rule-2}: Only segments longer than the tolerance $\tau$ are considered to reflect the overall shape of the polygon.
\item \emph{Rule-3}: Consistent simplifications should be performed for feature types such as intrusion/extrusion, offset, and corner.
\end{itemize}\vspace{-0.4cm}

\begin{figure}[h]
    \centering
    \includegraphics[width=\textwidth]{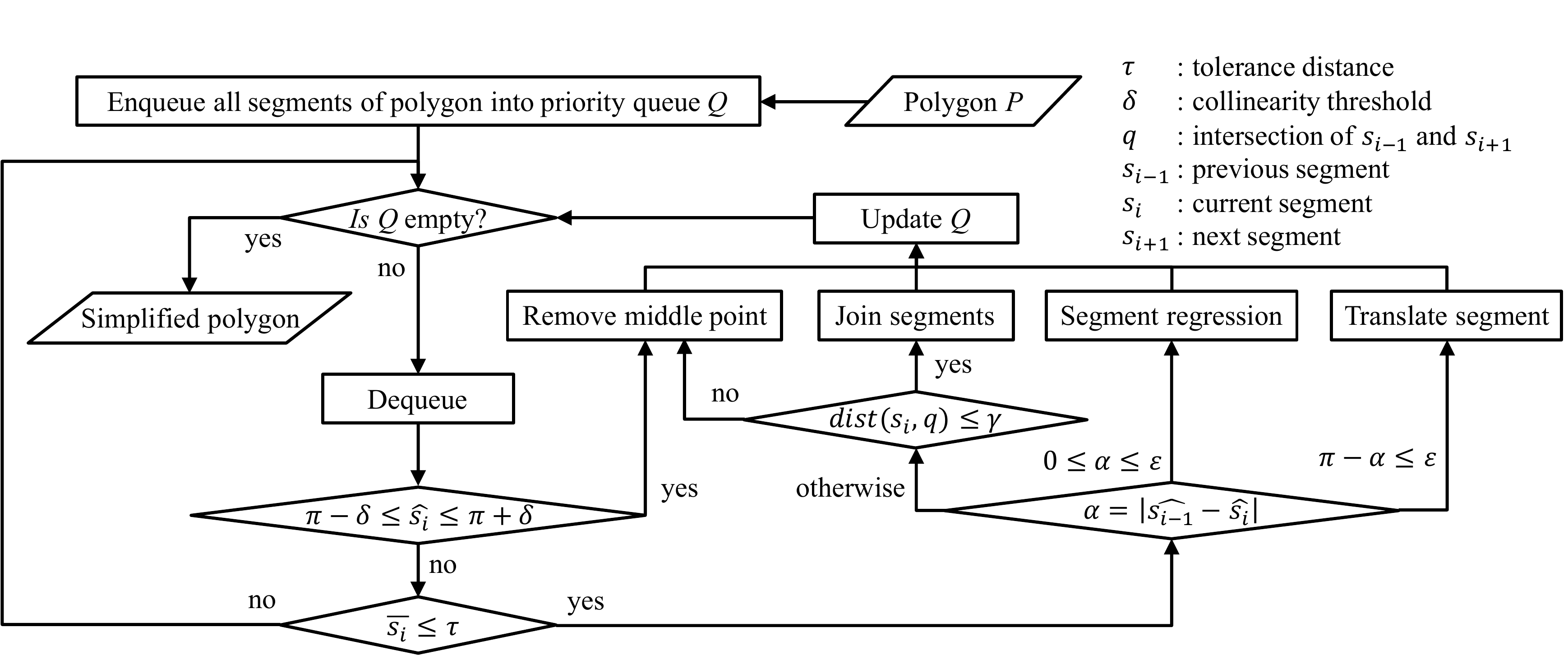}
    \caption{Flow chart of the algorithm of simplification of 2D polygon}
    \label{fig:footprint-simplification-flow}
\end{figure}

\begin{figure}[!h]
    \centering
    \includegraphics[width=16.5cm]{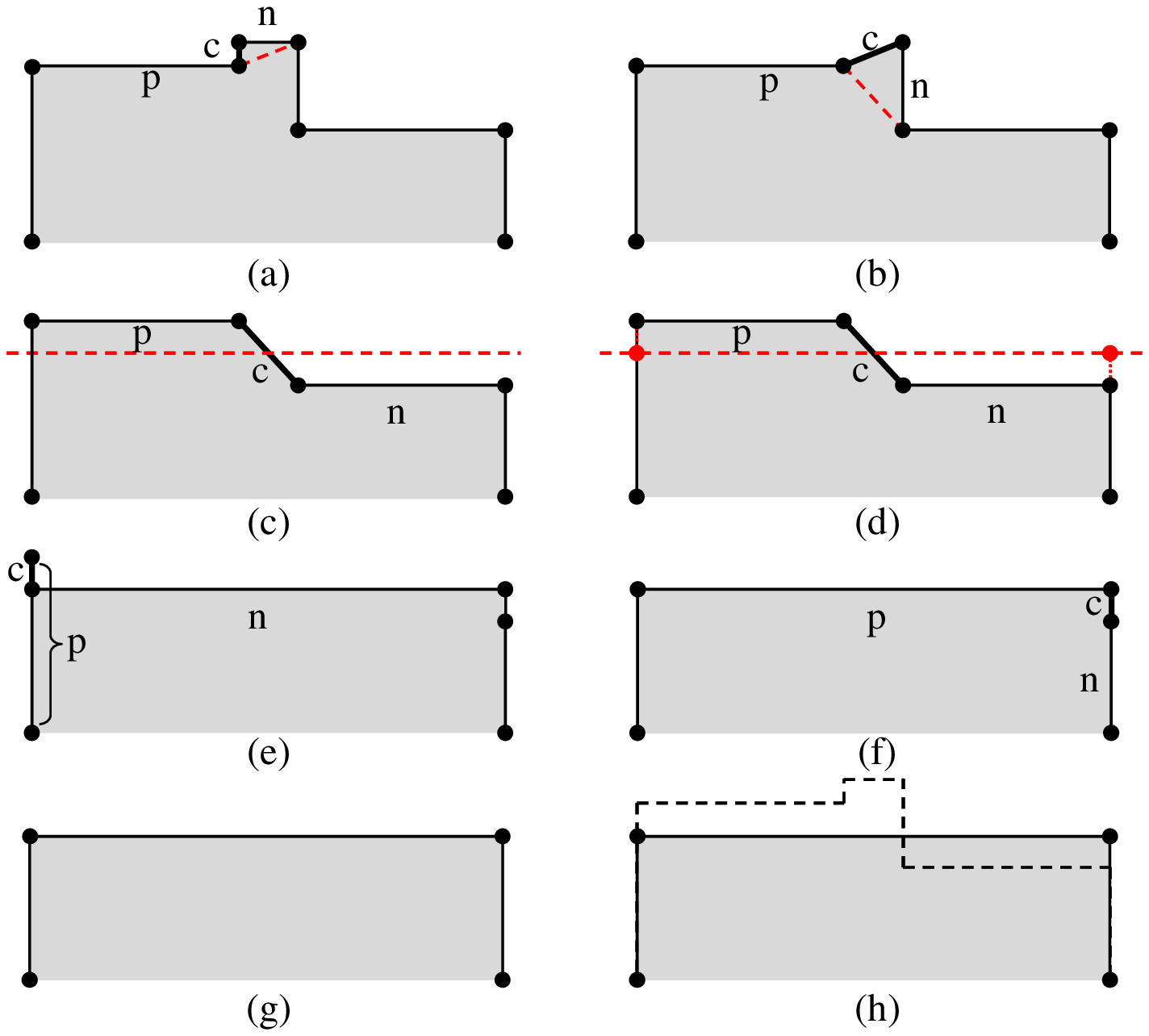}
    \caption{Removing points and merging into regression segment\vspace{-0.2cm}}
    \label{fig:removing-merging}
\end{figure}

\begin{figure}[!h]
    \centering
    \includegraphics[width=15cm]{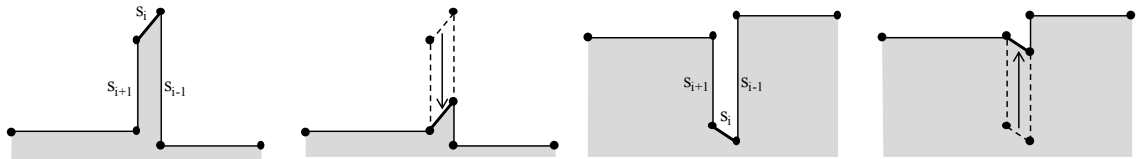}
    \caption{Translating current segment\vspace{-0.2cm}}
    \label{fig:translating}
\end{figure}

\begin{figure}[!h]
    \centering
    \includegraphics[width=8cm]{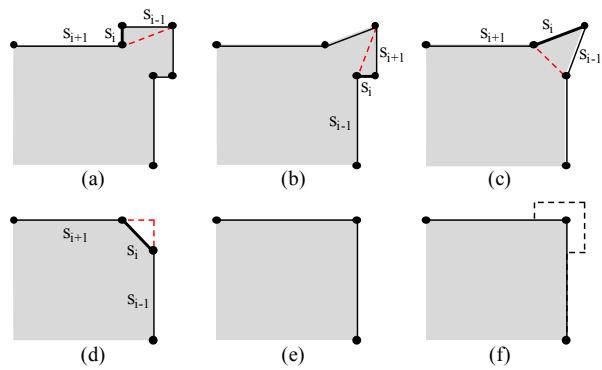}
    \caption{Removing of points and joining segments\vspace{-0.2cm}}
    \label{fig:removing-jointing}
\end{figure}

Figure \ref{fig:footprint-simplification-flow} depicts the flow of the simplification algorithm\footnote{The flow chart of \cite{kim2019simplification} was modified according to the notation and context.} that considers the rules, and Algorithm \ref{algo:Simplification} shows the corresponding pseudo-code. In order to simplify shorter segments first in accordance with \emph{Rule-1}, all segments of $P$ are sorted by length and inserted into a priority queue $Q$ (lines 1-3). Simplification is repeatedly carried out on the next segment in the queue until the queue is empty (line 4). The shortest line segment is dequeued from the queue as the current segment $s_i$ (line 5). To merge collinear segments, it is checked whether $\pi - \delta < \widehat{s_i} < \pi + \delta$, where $\delta>0$ is a small angle threshold that is used to evaluate collinearity (line 7). If so, the middle point between $s_i$ and $s_{i+1}$ is removed so that a new line segment between the two end points is created (line 8; see also Algorithm~\ref{algo:RemoveMiddlePoint}). If the length of current segment is less than or equal to $\tau$, simplification is performed (lines 9-21). In consideration of Rule-3, one of four methods is chosen depending on the angle difference $\alpha=|\widehat{s_{i-1}}-\widehat{s_i}| \le \pi$ (lines 10-21). Figures \ref{fig:removing-merging}-\ref{fig:removing-jointing} illustrate this process.

\begin{algorithm}[t]
\SetKwInOut{Input}{input}
\SetKwInOut{Output}{output}
\DontPrintSemicolon
\Input{$P$ \tcp*{Polygon to simplify}}
\Input{$\tau$ \tcp*{Tolerance distance}}
\Input{$\varepsilon$ \tcp*{Tolerance angle}}
\Input{$\delta$ \tcp*{Angle threshold used to determine if consecutive segments are collinear}}
\Input{$\gamma$ \tcp*{Distance threshold used to determine whether to join neighboring segments}}
\PrintSemicolon
\BlankLine
$Q \leftarrow \emptyset$ \tcp*{Initialize a priority queue}
\ForEach{$s \in P$} {
    $Q \leftarrow Q \cup s$ \tcp*{Insert all segments of polygon $P$ into the queue $Q$}
}
\While{$|Q| \ge 0$ {\bf and} $|P| \ge 3$}{
	$s_i \leftarrow Q.dequeue()$ \tcp*{Dequeue next segment}
	$\alpha \leftarrow |\widehat{s_{i-1}}-\widehat{s_i}|$ \tcp*{Angle difference between the angles entering and leaving $s_i$}
	\uIf{$\pi - \delta < \widehat{s_i} < \pi + \delta$}
    {
        $RemoveMiddlePoint(s_{i})$ \tcp*{Merge two approximately collinear consecutive segments}
    }
	\ElseIf{$\overline{s_i} \leq \tau$}
	{
    	\uIf{$0 \leq \alpha \leq \varepsilon $} 
    	{
        	$SegmentRegression(s_{i})$\;
    	}
    	\uElseIf{$\pi - \alpha \leq \varepsilon $} 
    	{
    	    $TranslateSegment(s_{i})$\;
    	}
    	\Else 
    	{
    	    $q \leftarrow Intersect(s_{i-1},s_{i+1})$  \tcp*{Intersection of two lines obtained by extending $s_{i-1}$ and $s_{i+1}$}
    	    \uIf{$dist(s_i,q) \leq \gamma$} 
    	    {
    	        $JoinSegment(s_{i},q)$\;
    	    }
    	    \uElseIf{$\overline{s_{i-1}} < \overline{s_{i+1}}$}
    	    {
    	       $RemoveMiddlePoint(s_{i-1})$\;
    	    }
    	    \Else
    	    {
    	        $RemoveMiddlePoint(s_{i})$\;
    	    }
    	}
	}
	
}
\Return $P$ \tcp*{Return the simplified polygon}

\caption{Simplification}
\label{algo:Simplification}
\end{algorithm}

\begin{figure}[!h]
    \centering
    \includegraphics[width=13cm]{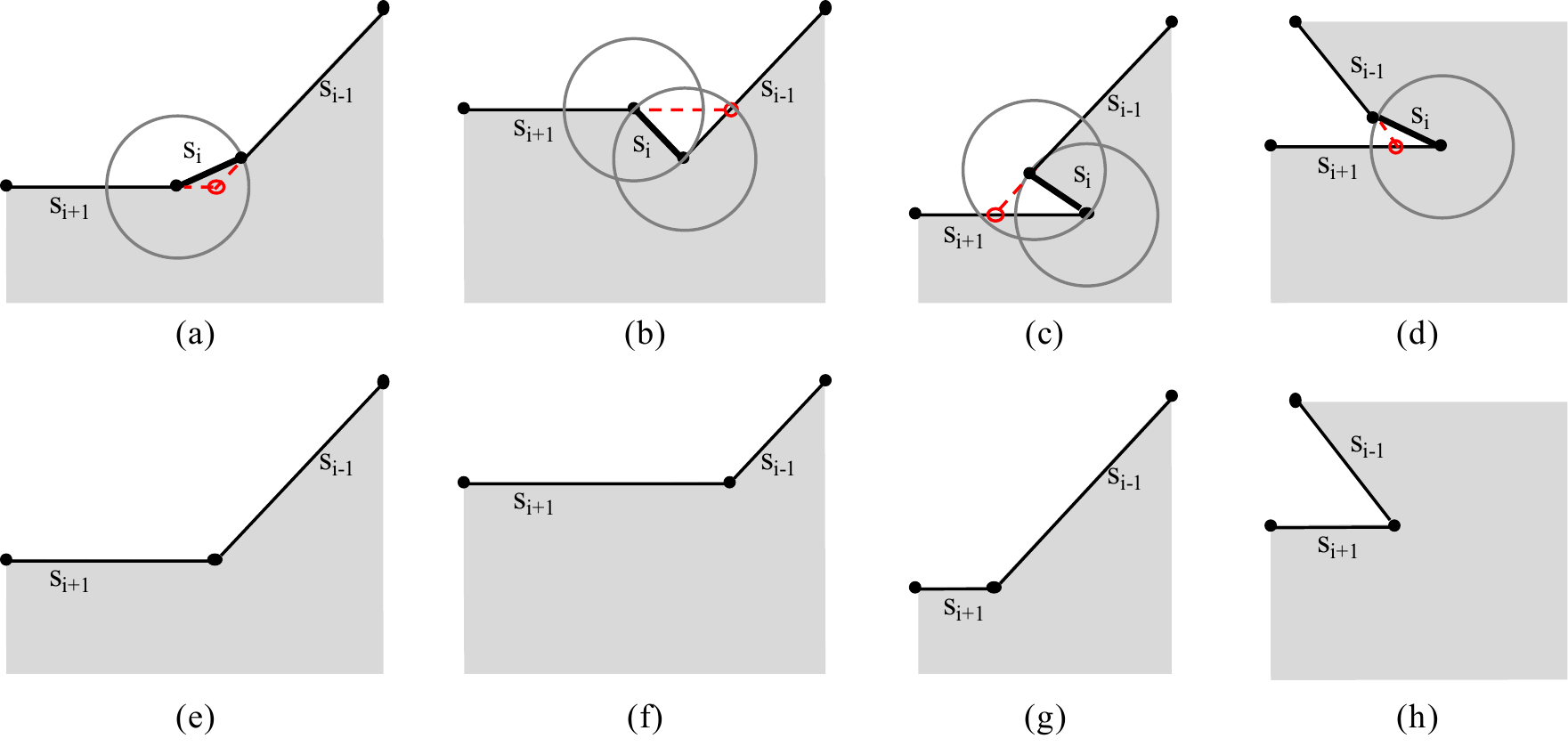}
    \caption{Joining segments ($\gamma = \overline{s_i}$)\vspace{-0.2cm}}
    \label{fig:joining}
\end{figure}

\begin{figure}[!h]
    \centering
    \includegraphics[width=16cm]{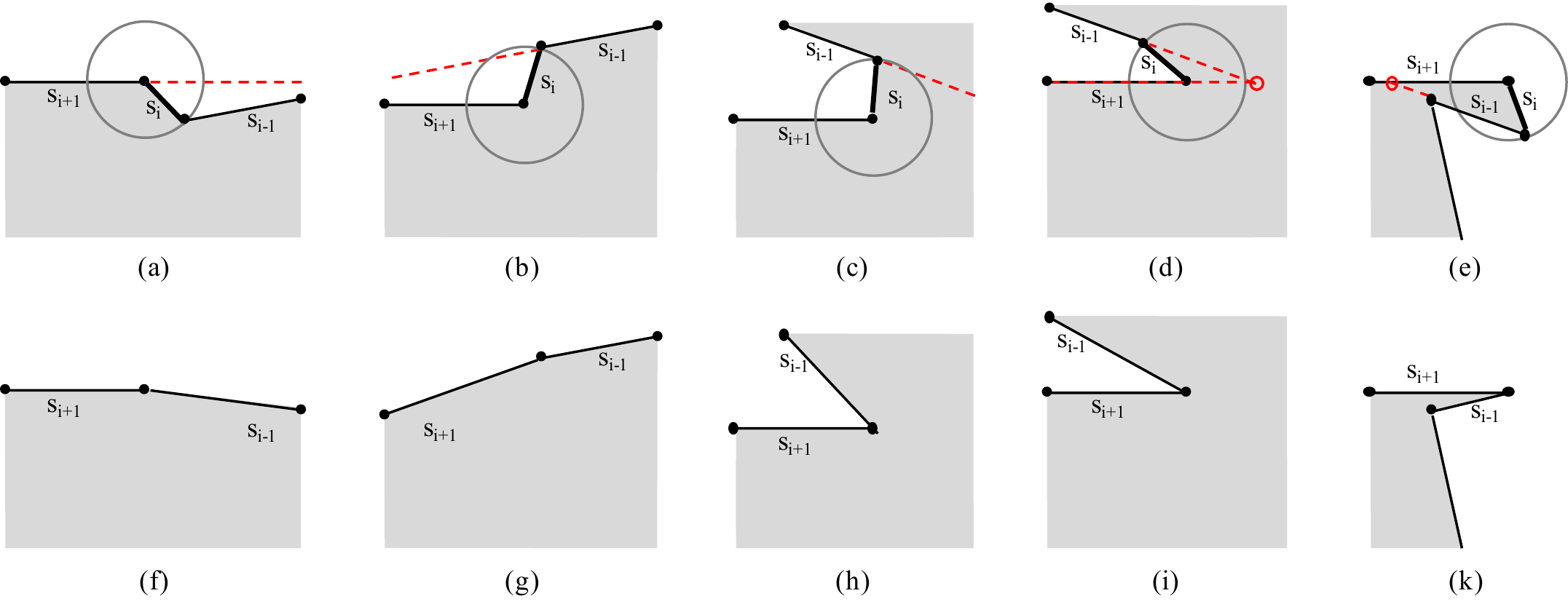}
    \caption{Removing middle points ($\gamma = \overline{s_i}$)\vspace{-0.2cm}}
    \label{fig:joining-fail}
\end{figure}

\begin{itemize}
\item If $0 \leq \alpha \leq \varepsilon$, regress two segments $s_{i-1}$ and $s_{i+1}$, considering their lengths and tangents as shown in Figures \ref{fig:removing-merging}(c), (d) and (e). The detailed process is outlined in Algorithm~\ref{algo:SegmentRegression}.
\item If $\pi - \alpha \leq \varepsilon$, this is considered an intrusion/extrusion, and the current segment is translated in order to remove the intrusion/extrusion as shown in Figure \ref{fig:translating} (see Algorithm~\ref{algo:TranslateSegment}).
\item Otherwise, it is checked whether the two segments can be joined by extending $s_{i-1}$ and $s_{i+1}$ until $s_{i-1}$ and $s_{i+1}$ intersect (see Figure \ref{fig:joining}). Let $q$ be the intersection point, $dist$ a distance function, and $\gamma$ the distance threshold.
\begin{itemize}
	\item If $dist(s_i,q) \le \gamma$ as shown in Figures \ref{fig:joining}(a)-(d), then remove $s_i$ and join $s_{i-1}$ and $s_{i+1}$ (see Algorithm~\ref{algo:JoinSegment} and Figure \ref{fig:joining}(e)-(h)). 
	\item Otherwise, do not join the segments because the extended part of the segments is too long as shown in Figures \ref{fig:joining-fail}(a)-(e). Instead, remove the middle point between $s_i$ and $s_{i-1}$ (or $s_{i+1}$) (see Figure  \ref{fig:joining-fail}(f)-(k)).
\end{itemize}
\end{itemize}

Pseudocode for the functions invoked in Algorithm~\ref{algo:Simplification} is described as Algorithms~\ref{algo:RemoveMiddlePoint}, \ref{algo:SegmentRegression}, \ref{algo:TranslateSegment}, and \ref{algo:JoinSegment}. Assume that $P$ and $Q$ defined in Algorithm~\ref{algo:Simplification} can be accessed from Algorithms~\ref{algo:RemoveMiddlePoint}, \ref{algo:SegmentRegression}, \ref{algo:TranslateSegment}, and \ref{algo:JoinSegment}. Python implementation and examples can be found at the git repository (\url{https://github.com/joonseok-kim/simplification}).

\begin{algorithm}
\SetKwInOut{Input}{input}
\SetKwInOut{Output}{output}
\DontPrintSemicolon
\Input{$s_k$ \tcp*{Segment}}
\PrintSemicolon
\BlankLine
$Q \leftarrow Q \setminus s_{k+1}$ \tcp*{Remove $s_{k+1}$ from the queue $Q$}
$s' \leftarrow (p_{k},p_{k+2})$ \tcp*{Create a new segment ($p_{k},p_{k+2}$) by merging $s_k$ and $s_{k+1}$}
$P \leftarrow P \setminus (s_k \cup s_{k+1})$ \tcp*{Remove existing $s_k$ and $s_{k+1}$ from $P$}
$Q \leftarrow Q \cup s'$ \tcp*{Insert the new segment into the queue $Q$}
\caption{RemoveMiddlePoint}
\label{algo:RemoveMiddlePoint}
\end{algorithm}

\begin{algorithm}
\SetKwInOut{Input}{input}
\SetKwInOut{Output}{output}
\DontPrintSemicolon
\Input{$s_k$ \tcp*{Segment}}
\PrintSemicolon
\BlankLine
$Q \leftarrow Q \setminus (s_{k-1} \cup  s_{k+1})$\;
$r \leftarrow \overline{s_{k-1}} / (\overline{s_{k-1}} + \overline{s_{k+1}})$\ \tcp*{Ratio to be used as a weight.}
$p \leftarrow s_k.PointAlong(r)$  \tcp*{A point $p$ on $s_k$ such that $p$ is at distance $\overline{s_k}\cdot r$ from $p_k$.}
$\theta \leftarrow$ tan$(\widehat{s_{k-1}} \cdot r + \widehat{s_{k+1}} \cdot (1-r))$ \tcp*{The slope of a regression line for $s_{k-1}$ and $s_{k+1}$.}
$q_1 \leftarrow Projection(s_{k-2},p,\theta)$ \tcp*{Intersection of $s_{k-2}$ with the line through $p$ with slope $\theta$.}
$q_2 \leftarrow Projection(s_{k+2},p,\theta)$ \tcp*{Intersection of $s_{k+2}$ with the line through $p$ with slope $\theta$.}
$s_k \leftarrow (q_1,q_2)$  \tcp*{A regression line segment for $s_{k-1}$ and $s_{k+1}$.}
$s_{k-1} \leftarrow (p_{k-1},q_1)$ \tcp*{Update $s_{k-1}$}
$s_{k+1} \leftarrow (q_2,p_{k+2})$ \tcp*{Update $s_{k+1}$}
$Q \leftarrow Q \cup s_{k-1} \cup s_{k} \cup s_{k+1}$ \tcp*{Add the new three segments into the queue}
\caption{SegmentRegression}
\label{algo:SegmentRegression}
\end{algorithm}

\begin{algorithm}
\SetKwInOut{Input}{input}
\SetKwInOut{Output}{output}
\DontPrintSemicolon
\Input{$s_k$ \tcp*{Segment}}
\PrintSemicolon
\BlankLine
$Q \leftarrow Q \setminus (s_{k-1} \cup s_{k+1})$ \tcp*{Remove $s_{k-1}$ and $s_{k+1}$ from the queue $Q$}
\uIf{$\overline{s_{k-1}} < \overline{s_{k+1}}$}
{
    $p' \leftarrow \overrightarrow{p_{k+1}}-\overrightarrow{s_{k-1}}$ \tcp*{Translate vertex $p_{k+1}$ by vector $s_{k-1}$}
    $s_k \leftarrow (p_{k-1},p')$ \tcp*{Update $s_k$}
    $s_{k+1} \leftarrow (p',p_{k+2})$ \tcp*{Update $s_{k+1}$}
    $P \leftarrow P \setminus s_{k-1}$ \tcp*{Remove existing $s_{k-1}$}
    $Q \leftarrow Q \cup (s_k \cup s_{k+1})$ \tcp*{Add $s_k$ and $s_{k+1}$ into $Q$}
}
\uElseIf{$\overline{s_{k-1}} > \overline{s_{k+1}}$}
{
    $p' \leftarrow \overrightarrow{p_{k}}-\overrightarrow{s_{k+1}}$ \tcp*{Translate vertex $p_{k}$ by vector $s_{k+1}$}
    $s_{k-1} \leftarrow (p_{k-1},p')$ \tcp*{Update $s_{k-1}$}
    $s_k \leftarrow (p',p_{k+2})$ \tcp*{Update $s_k$}
    $P \leftarrow P \setminus s_{k+1}$ \tcp*{Remove existing $s_{k+1}$}
    $Q \leftarrow Q \cup (s_{k-1} \cup s_k)$ \tcp*{Add $s_{k-1}$ and $s_k$ into $Q$}
}
\Else
{
    $s_k \leftarrow (p_{k-1},p_{k+2})$ \tcp*{Update $s_k$}
    $P \leftarrow P \setminus (s_{k-1} \cup s_{k+1})$ \tcp*{Remove existing $s_{k-1}$ and $s_{k+1}$}
    $Q \leftarrow Q \cup s_k$ \tcp*{Add $s_k$ into $Q$}
}
\caption{TranslateSegment}
\label{algo:TranslateSegment}
\end{algorithm}

\begin{algorithm}
\SetKwInOut{Input}{input}
\SetKwInOut{Output}{output}
\DontPrintSemicolon
\Input{$s_k$ \tcp*{Segment}}
\Input{$q$ \tcp*{Intersection point}}
\PrintSemicolon
\BlankLine
$Q \leftarrow Q \setminus (s_{k-1} \cup s_{k+1})$ \tcp*{Remove $s_{k-1}$ and $s_{k+1}$ from the queue $Q$}
$s_{k-1} \leftarrow (p_{k-1},q)$ \tcp*{Update $s_{k-1}$}
$s_{k+1} \leftarrow (q,p_{k+2})$ \tcp*{Update $s_{k+1}$}
$P \leftarrow P \setminus s_k$ \tcp*{Remove existing $s_k$}
$Q \leftarrow Q \cup (s_{k-1} \cup s_{k+1})$ \tcp*{Add $s_{k-1}$ and  $s_{k+1}$ into $Q$}
\caption{JoinSegment}
\label{algo:JoinSegment}
\end{algorithm}

\section{Experiments}

\begin{figure}[!h]
    \centering
    \begin{minipage}[t]{.5\textwidth}
        \includegraphics[width=\textwidth]{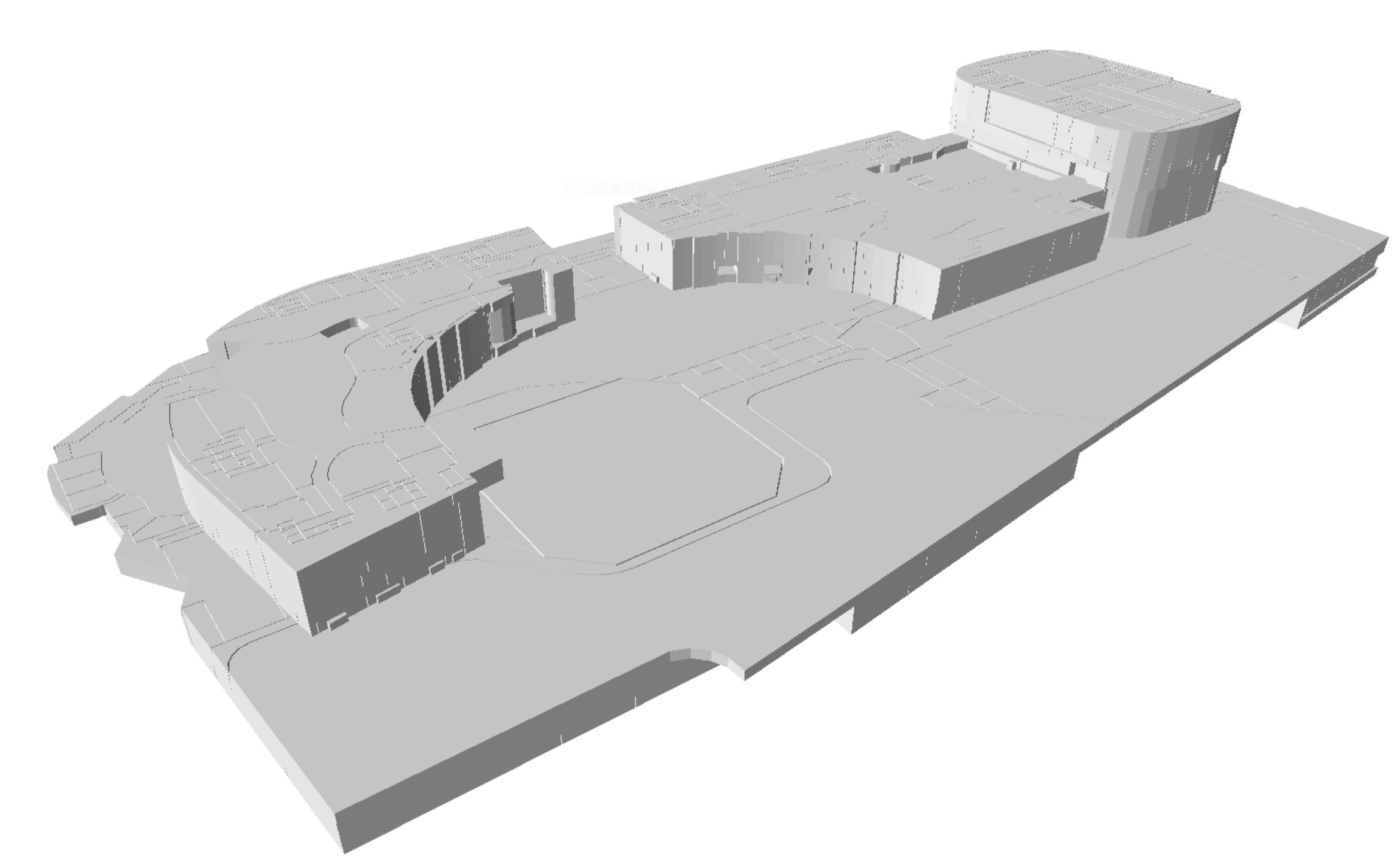}
        \caption{Lotte World Mall dataset}
        \label{fig:dataset-lotte}
    \end{minipage}%
    \begin{minipage}[t]{.5\textwidth}
        \includegraphics[width=\textwidth]{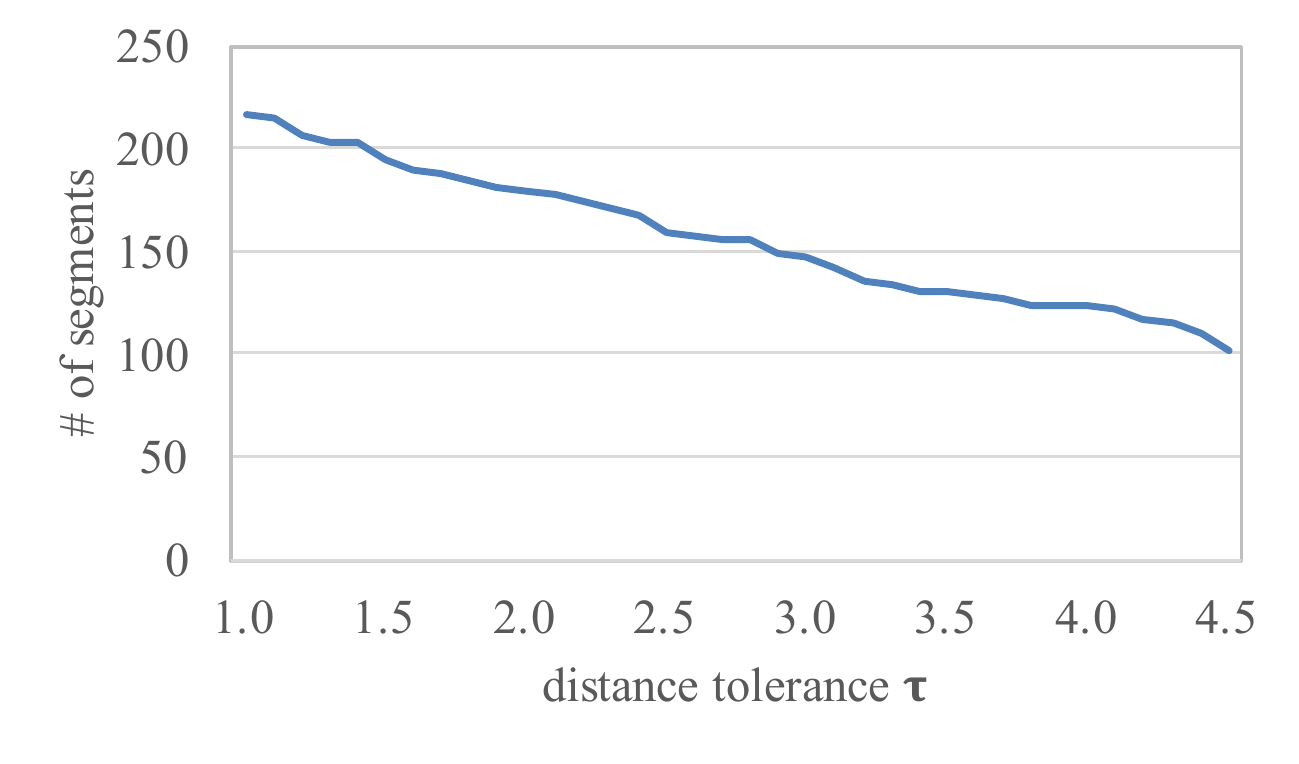}
        \caption{The number of segments left after simplification of the corridor shown in Figure \ref{fig:realdata}}
        \label{fig:experiment}
    \end{minipage}
\end{figure}

\begin{figure}[t]
	\centering
    \subfloat[initial\label{fig:intial}]{%
       \includegraphics[width=0.245\textwidth]{fig/initial.png}
    }
    \subfloat[$\tau=1.5$\label{fig:1-5}]{%
       \includegraphics[width=0.245\textwidth]{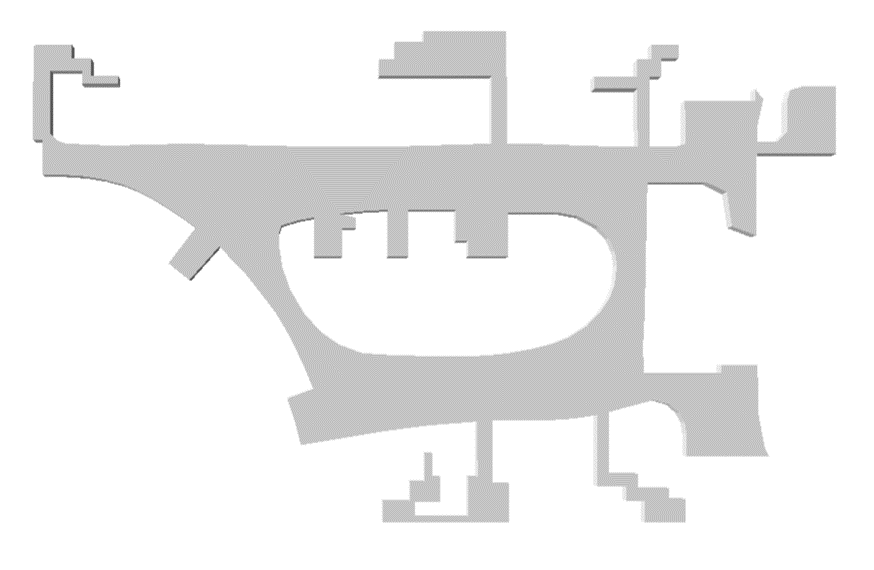}
    }
    \subfloat[$\tau=2$\label{fig:2}]{%
       \includegraphics[width=0.245\textwidth]{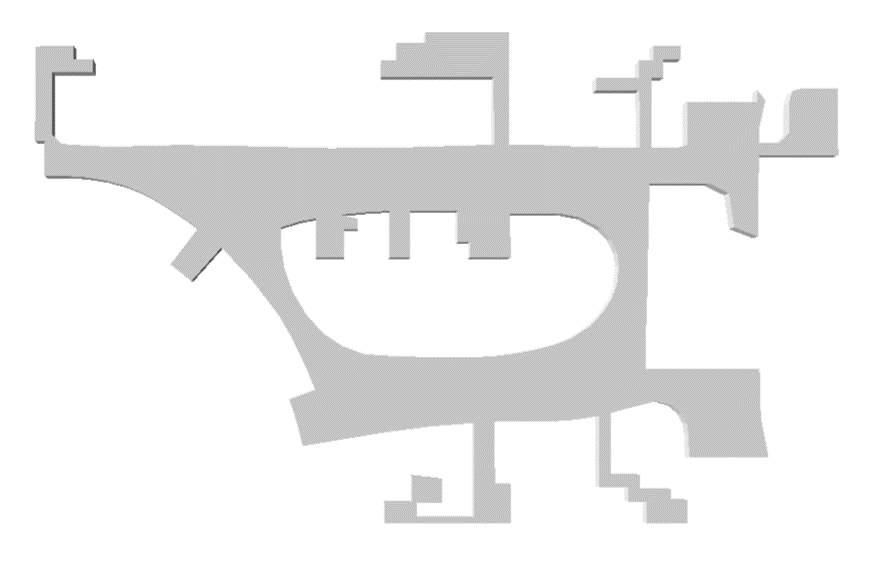}
    }
    \subfloat[$\tau=2.5$\label{fig:2-5}]{%
       \includegraphics[width=0.245\textwidth]{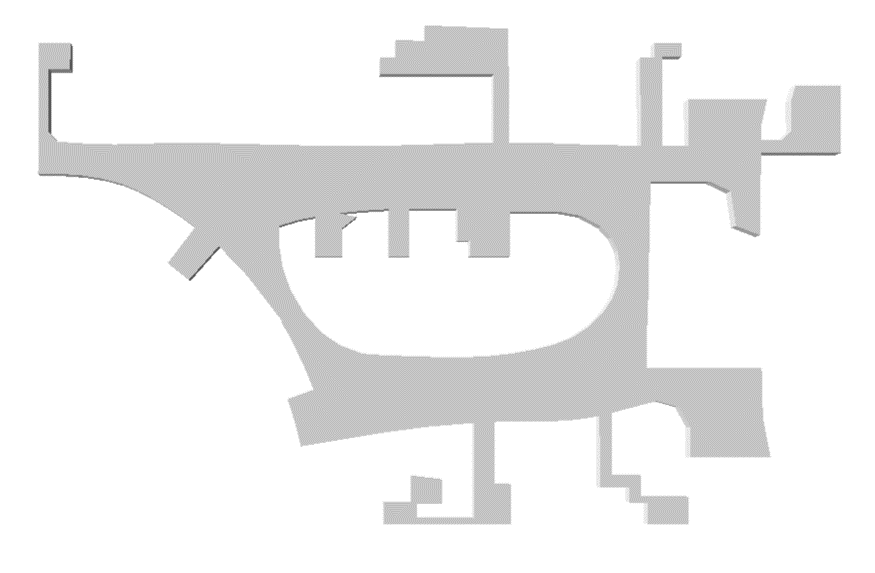}
    }
    \\
    \subfloat[$\tau=3$\label{fig:3}]{%
       \includegraphics[width=0.245\textwidth]{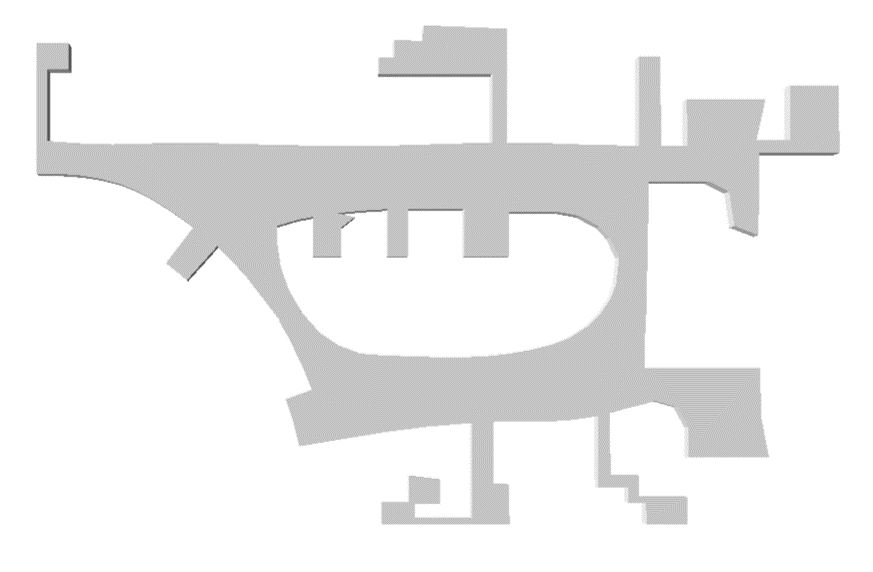}
    }
    \subfloat[$\tau=3.5$\label{fig:3-5}]{%
       \includegraphics[width=0.245\textwidth]{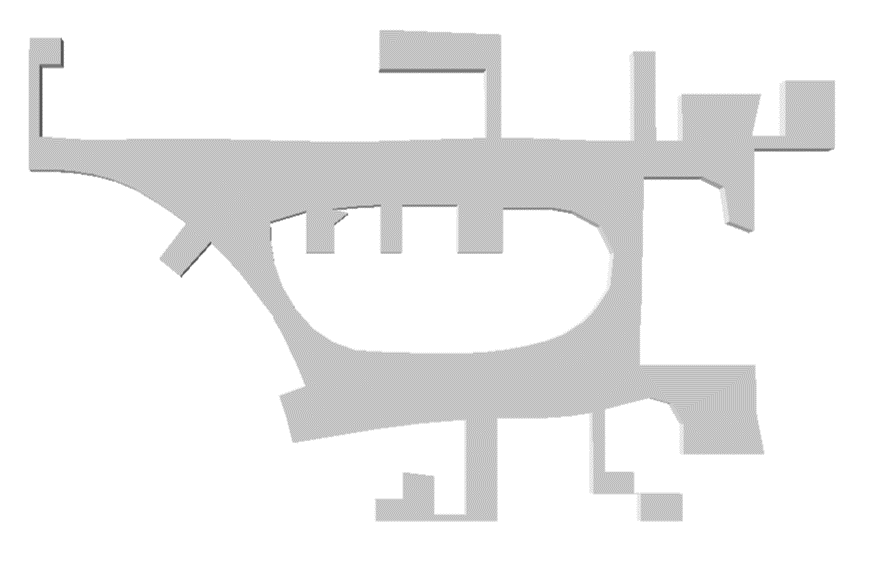}
    }
    \subfloat[$\tau=4$\label{fig:4}]{%
       \includegraphics[width=0.245\textwidth]{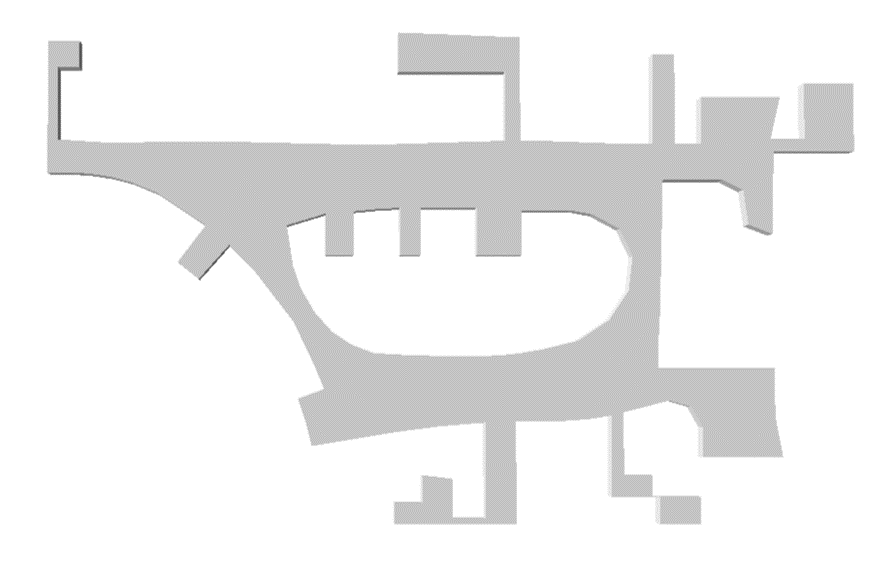}
    }
    \subfloat[$\tau=4.5$\label{fig:4-5}]{%
       \includegraphics[width=0.245\textwidth]{fig/4-5.png}
    }
    \caption{Results of simplification for varying $\tau$\vspace{-0.2cm}}
    \label{fig:realdata}
\end{figure}

IndoorGML data for the  Lotte World Mall\footnote{IndoorGML data (core module) for Lotte World Mall (IndoorGML 1.0.3), \url{http://www.indoorgml.net/resources/}} (LWM), see Figure \ref{fig:dataset-lotte}, one of the most complex shopping malls in Seoul, is used to conduct a comparative experiment on the performance of the introduced simplification. IndoorGML is one of OGC standards to provide a standard framework of semantic, topological, and geometric models for indoor spatial information \cite{kang2017}. The dataset for the experiment is compatible with CityGML LoD4 \cite{kim2014} so that data can be visualzed via any CityGML viewer. Figures \ref{fig:experiment} and \ref{fig:realdata} show the visual and quantitative results of simplification of one large and complex corridor for varying $\tau$, given $\varepsilon = \pi/36, \delta = \pi/180, \gamma = \overline{s_i}$. As $\tau$ increases, gradual simplification of shorter segments is observed, in particular at extremities,  dominant features such as intrusions, extrusions, offsets, and corners are preserved.

\begin{figure}[!h]
    \centering
    \includegraphics[width=\textwidth]{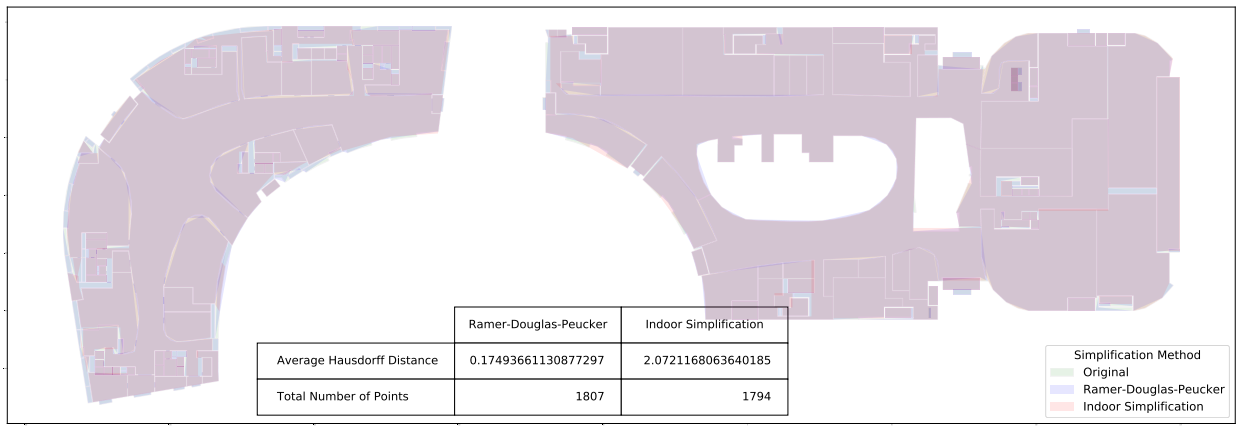}
    \caption{Comparison between the original data, the indoor simplification ($\tau$=2), and RDP (tolerance=1.2)}
    \label{fig:result-lwm}
\end{figure}

While Figure \ref{fig:realdata} focuses on simplification results for one polygon, Figure \ref{fig:result-lwm} shows the comparison between the original LWM data, the indoor simplification, and RDP for all spaces on a floor. Note that a needle-shaped polygon will vanish after simplification if the length of its width is less than $\tau$.

\bibliographystyle{unsrt}

\bibliography{main}

\end{document}